# Towards Efficient Machine Learning Method for IoT DDoS Attack Detection

Pavitra Modi

Faculty of Computer Science

*University of New Brunswick, Fredericton, NB, Canada*

pmodi@unb.ca

*Abstract*— **With the rise in the number of IoT devices and its users, security in IoT has become a big concern to ensure the protection from harmful security attacks. In the recent years, different variants of DDoS attacks have been on the rise in IoT devices. Failure to detect DDoS attacks at the right time can result in financial and reputational loss for victim organizations. These attacks conducted with IoT devices can cause a significant downtime of applications running on the Internet. Although researchers have developed and utilized specialized models using artificial intelligence techniques, these models do not provide the best accuracy as there is always a scope of improvement until 100% accuracy is attained. We propose a hybrid feature selection algorithm that selects only the most useful features and passes those features into an XGBoost model, the results of which are explained using feature importances. Our model attains an accuracy of 99.993% on the CIC IDS 2017 dataset and a recall of 97.64 % on the CIC IoT 2023 dataset. Overall, this research would help researchers and implementers in the field of detecting IoT DDoS attacks by providing a more accurate and comparable model.**

## I. INTRODUCTION

In the contemporary world, we are surrounded by IoT devices. Be it home appliances, smartphones, laptops, or cameras, IoT devices form an inseparable part of our lives. According to statista, the total number of IoT devices has grown from approximately 15 billion in 2015 to approximately 62 billion in 2024 and it is expected to rise to approximately 75 billion in 2025 [1].

To make IoT devices more user friendly, manufacturers of IoT devices make people add their personal information within these devices resulting in more convenience for the users of IoT devices but raising a huge security threat for the users [2], [3]. To make the situation worse for users security-wise, these manufacturers make IoT devices using cheap components, which do not have strong security protections in place [2], [3]. Along with this, a loop hole that results in IoT devices becoming a victim of DDoS attacks is that these devices do not receive frequent security updates [3]. Moreover, as these devices are small, it makes it very hard for the users to check for any malicious activities in these devices [3]. IoT devices become victims of cybersecurity attacks mainly due to the wrong use of protocol by the IoT device [4].

The responsibility to protect the IoT devices from security threats not only rests in the hands of the users but also rests in the hands of the manufacturers [3]. This is because although the users can follow best security practices to use IoT devices, manufacturers need to provide frequent security updates for these IoT devices and should create the IoT device with great security practices like secure password management [3].

There are many different types of cybersecurity attacks possible on IoT devices [2]. DDoS attacks, in particular, occurs when a large number of packets are sent to the server through various devices, which results in the server not being able to respond to the packets of genuine devices [5]. A critical issue that arises with these type of attacks is that DDoS attacks can result in a server outage that ranges from some hours to a few days [4]. This, ultimately, causes a huge dent on the reputation of the victim organization, which may cause a significant reduction in the faith of the client in the organization [4]. For example, Kaspersky [6] reported that the year 2018 recorded one of the longest, in terms of time consumed to perform the attack, DdoS attacks that lasted for close to greater than 12 days. Moreover, Osborne [7] from zdnet reported that if DdoS attacks hit an organization, then it would cost the organization approximately 2.5 million dollars, which highlighted the criticality of these attacks.

One of the big names in the world of IoT DdoS attacks is Mirai botnet [24]. As reported by Flashpoint and Williams [8], Mirai malware starts by creating a botnet of a huge number of various kinds of IoT devices, which are then used to perform a DdoS attack on a target. Mirai malware showed its potential and power by creating a botnet of IoT devices and clogging the servers of a company named Dyn [8]. This, ultimately, resulted in numerous websites becoming unavailable to the users for some period of time [8]. This attack stresses the importance of detecting DdoS attacks in IoT devices [4].

To protect against these attacks, machine learning based classification methods have shown great progress in both normal devices and IoT devices [4]. Machine learning methods are considered great in this scenario because they have been observed to classify a vast number of packets as benign or malicions [4]. Not only machine learning methods, but also deep learning [9] methods have also proven to give great results towards detecting DdoS attacks in IoT devices [10].

In this paper, we aim to use machine learning methods to solve of problem of detection of DdoS attacks in IoT devices. To accomplish this goal, we have developed a hybrid feature selection algorithm, which is used to select optimal number of features in the datasets to achieve great accuracy, precision, recall, and F1-Score on machine learning models. In comparison to the other models, we have made the models explainable and interpretable. By doing so, we have explained which features in Intrusion Detection Systems of normal devices and IoT systems contribute the most towards the detection of DdoS attacks in these systems. To distinguish between the most important features used for detecting DdoS attacks in Intrusion Detection Systems of normal devices and IoT devices, we have considered datasets of both the types of devices. The research contributions of this paper are as follows:

- We have proposed a hybrid feature selection algorithm used to perform feature selection on datasets.
- We developed a machine learning model combined with the proposed feature selection algorithm, which resulted in the creation of a superior model for detecting DdoS attacks in normal devices and a comparable model to detecting DdoS attacks in IoT devices.
- We made both of our models explainable and interpretable by presenting the set of features that prove to be helpful in detecting DdoS attacks in normal and IoT devices.

The remainder of this paper is organized as follows. We provide a detailed review of the works in the field of detecting DdoS attacks in IoT devices in section II. Following that, we describe in detail the process of conduct the research including but not limited to the proposed feature selection algorithm in section III. Section IV discusses the processes and architecture used to train the models. Section V highlights the results and discusses the key findings of the research. Finally, we end the research paper by the conclusion and discussing the future works of this research.

## II. Related Works

In this section, we will present research done in the field of IoT DdoS attack detection using artificial intelligence techniques. We have organized this section in such a way that we cover most if not all of the techniques used until now to detect DdoS attacks in IoT devices.

To solve the problem of detecting DdoS attacks in IoT devices, machine learning and deep learning models have been implemented in previous research papers [10]. Chen et al. [11] proposed a model that spans over multiple layers in a system to detect DdoS attacks. The machine learning model used for this approach generated a F1-score of more than 97% [11]. Chen et al. [11] suggested that unsupervised machine learning techniques have the potential to provide an improvement to their existing model [11].

Doshi et al. [12] created their own IoT DdoS dataset in their local environment. Doshi et al. [12] used this dataset to train machine learing and deep learning [9] models, and selected the best model as their proposed model. Doshi et al. [12] found that their Random Forest Model [13] performed the best with an accuracy of 0.999. Their dataset consisted of DdoS packets that were more than ten times the total number of benign packets [12]. This acted as a central limitation towards the reseults they had achieved [12].

Aysa et al. [14] created their own IoT DdoS dataset and used this dataset to train different types of machine learning models . They performed feature selection using solely the Pearson correlation coefficient [15]. Using this feature selection technique, they selected 40 features, and passed those features into different models to compare these models on various metrics [14]. They found that the combination of random forest [13] and decision trees [16] produced the best results.

Wang et al. [17] used datasets like CIC-IDS 2017 [18] and CIC IoT 2023 [19] to create a deep learning [9] based model for detecting IoT DdoS attacks in a less memory intensive method. Wang et al. [36] proposed a combination of a deep neural network [20] and a Bidirectional Long Short Term Memory [21] model. This model attained an accuracy of 99.96 % [17].

Ma et al. [10] proposed a convolutional neural network [22] model with a modified loss function. Rather than developing their own dataset, they used the NSL KDD [23] dataset to perform their experiments and found that their model achieved an accuracy of 92.99 % [10]. Ma et al. [10] considered to further improve their network architecture to classsify different types of DdoS attacks in IoT devices.

Nkoro et al. [24] proposed an explainable deep neural network [25] with the help of SHAP [26] values and other frameworks. This model used the 2023 EdgeIIoT dataset [27]. This model utilized a Filter Feature selection algorithm to find out the relevant features and used those features to train models [24]. Following that, Nkoro et al. [24] performed cross-validation [28] across 10 folds. Their model attained a F1-score of 99.9% [24].

Ahmim et al. [29] created a model using a combination of deep learning [9] models with the aim of detecting DDoS attacks in IoT devices. Their proposed model attained an accuracy of 80.75 % on the CIC-DDoS2019 [30] dataset [29]. Yuan et al. [31] used a sliding window in LSTM to train a model to solve the problem of detecting DDoS attacks by classifying sequences of packets rather than individual packets. Their model got a final accuracy of 97.606 % on the ISCX2012 [33] dataset [31].

Cvitić et al. [34] used logistic model trees to detech DdoS attacks in IoT devices. Cvitić et al. [34] used four datasets and to test the effectiveness of their model and found that their model attained an accuracy value in the range of 0.9921 to 0.9996.

## III. Research Methods.

*A. Dataset Selection*

For the dataset selection step, the CIC IoT 2023 [19] dataset is chosen. This dataset is chosen because it is a more recent dataset with many different types of IoT DdoS attacks [19]. To try this analysis on a different dataset, the CIC IDS 2017 [18] dataset is also chosen. This is mainly done to check the accuracy of the proposed feature selection algorithm on both the datasets and see if the accuracy of the models trained on both these datasets are better than the state-of-the-art models in the fields of cybersecurity and artificial intelligence. Moreover, by selecting these two datasets our aim would be to compare the features of both the datasets that provide a very high contribution to predict the results.

### B. Data Preparation and Preprocessing

#### B.A.1   For CIC IoT 2023 [19] dataset

This dataset did not have any null values, so data cleaning was not required. As the goal of this research project is to detect DdoS/DoS and benign packets, the "label" column was transformed to have a 0 represent "BenignTraffic" and a 1 for any type of DoS or DdoS packet.

#### B.A.2   For CIC IDS 2017 [18] dataset

As part of the Data Preparation and Preprocessing step, the dataset was cleaned to remove the packets with 'NaN' and 'infinity' values within the cells of the CICIDS 2017 Dataset. This process ensured that the dataset was cleaned and was ready for the next step of feature scaling.

### C. Feature Scaling

#### C.A.1   For CIC IoT 2023 [19] dataset

For the feature scaling step on this dataset, Standard scaler was used to scale the values in the dataset. Bacause after the Dataset preparation and preprocessing step, there were no text or categorical values, the standard scaling got done without any errors.

#### C.A.2   For CIC IDS 2017 [18] dataset

As part of the Feature Scaling Step, the non-categorical features of the dataset were scaled using Standard Scaler algorithm. Some categorical varibles like 'Flow ID', 'Source IP', and 'Destination IP' were removed from the dataframe before standard scaling was applied to ensure the Feature Scaling process went smoothly.

### D. Feature Selection

As part of this research, a new hybrid feature selection algorithm is being proposed. This hybrid feature selection algorithm is depicted as in the flowchart in Figure 1.

#### D.A.1   Proposed Hybrid Feature Selection Algorithm

The feature selection algorithm finds out the pearson correlation [15] coefficient between of all the features and the output label.

Let the set of all the features be represented by a matrix $F_{all}$, where $F_{all}$ represents all the features present in the dataset.

$$F_{all} = [f_1, f_2, \ldots, f_k]$$

Let the pearson correlation [15] coeffieienct values of each feature to the output label be represented by the matrix $P_{Correlation}$. Let $\mu_{pearson\_positive}$ represent the mean of positive pearson correlation [15] values and let $\mu_{pearson\_neagtive}$ represent the mean of the negative pearson correlation [15] values.

Let $a_1$ represent a set of features selected in the first step. Let $a_2$ represent the set of features not selected in the first step. Let $a_3$ represent the set of features selected in step 2. Let $a_4$ be the set that represents the set of feature obtained by performing the union of $a_1$ and $a_3$.

Let $\mu_{spearman\_positive}$ represent the mean of positive spearman correlation [35] values and let $\mu_{spearman\_neagtive}$ represent the mean of the negative spearman correlation [35] values. Let $\mu_{kendall\_positive}$ represent the mean of positive kendall correlation [36] values and let $\mu_{kendall\_neagtive}$ represent the mean of the negative kendall correlation [36] values.

Let $\mu_{spearman\_kendall\_positive}$ and $\mu_{spearman\_kendall\_neagtive}$ be calculated as follows:

$$\mu_{spearman\_kendall\_positive} = \frac{\mu_{kendall\_positive} + \mu_{spearman\_positive}}{2}$$

$$\mu_{spearman\_kendall\_negative} = \frac{\mu_{kendall\_negative} + \mu_{spearman\_negative}}{2}$$

Let $\mu_{information\_gain}$ represent the mean of the information gain values of each feature. Let $a_5$ be the set that includes all features selected by the information gain algorithm. Let $a_6$ represent the intersection of sets $a_4$ and $a_5$.

***Step 1:*** The feature selection algorithm finds out the pearson correlation [15] coefficient between of all the features and the output label. For the pearson correlation [15] coefficient's values found for each feature, the algorithm would calculate the values of $\mu_{pearson\_positive}$ and $\mu_{pearson\_neagtive}$.

Now, the algorithm would loop through the pearson [15] value of each feature and would add it to the set $a_1$ if the value is positive and it is greater than or equal to $\mu_{pearson\_positive}$ or if the pearson [15] value of the feature is negative and it is less equal to than the $\mu_{pearson\_neagtive}$. Once the set $a_1$ is generated, a new set is created named $a_2$. $a_2$ will include all features not selected in *Step 1*.

***Step 2:*** For the features in list $a_2$, spearman [35] and kendall tau [36] coefficients are found for each an every feature. Now, for the features in the set $a_2$, the values of $\mu_{spearman\_positive}$, $\mu_{spearman\_neagtive}$, $\mu_{kendall\_positive}$, $\mu_{kendall\_neagtive}$, $\mu_{spearman\_kendall\_positive}$, and $\mu_{spearman\_kendall\_negative}$ are found.

Now, the algorithm would loop through the average of the *spearman* [35] and kendall tau [36] values of each feature not selected in the first step would add it to the list $a_3$ if the average of spearman [35] and kendall tau [36] value for the feature is positive and it is greater than equal to $\mu_{spearman\_kendall\_positive}$ or if the average of speaman [35] and kendall tau [36] value for the feature of the feature is negative and it is less equal to than the $\mu_{spearman\_kendall\_negative}$. Once the set $a_3$ is formed, a union operation on both the set $a_1$ and $a_3$ and the output is stored in the set $a_4$. The features not selected in step 2 are removed.

*Step 3:* Now, on the set of all the original features before any modification, information gain [37] is applied to the features to rank the features based on their importance. The mean of the information gain values of the features is found. All features with the information gain value more than the mean value would be selected and added to list a5.

Following this step, a set intersection is found between the features selected by the correlation coefficients and information gain, that is, a set intersection is found between a4 and a5. The features found with this intersection are stored in a6. The features in a6 would be selected for the next step. This process is depicted in the Figure 1.

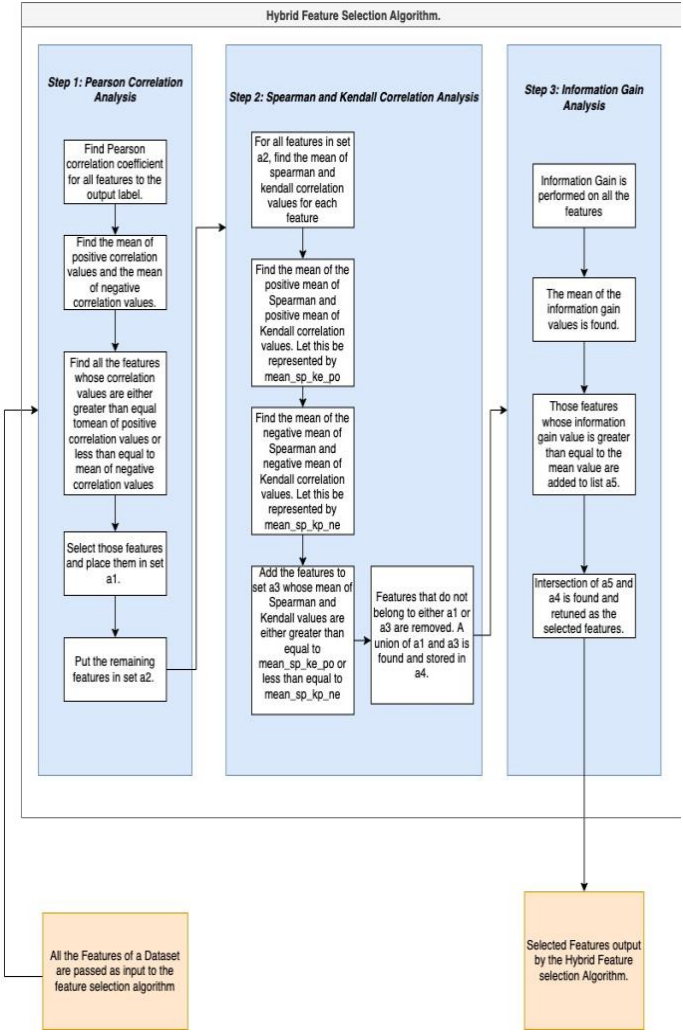

Fig. 1. Hybrid Feature Selection Algorithm

## IV. EXPERIMENTS AND DISCUSSION

### A. Computer Architecture for experiments

To perform the experiments, a macbook pro with 16 GB RAM architecture was used. The code for the experiements was written in Python programming languages with the help of modules and libraries.

### B. Metrics used for evaluation

To compare the results of this research paper with the other model, the evaluation metrics like Accuracy, Precision, Recall, F1-Score.

According to [38], accuracy represents the total number of packets that are correctly classified. Precision can be represented as the true positives divided by the sum of true positives ad false positives [38]. Recall represents the true positives divided by the sum of true positives and true negatives [38]. F1 Score resrpesents the mixture of recall and precision by giving a value that is calculated as the multiplication of precision and recall divided by the sum of precision and recall [38].

$$Accuracy = \frac{TP + TN}{TP + TN + FP + FN} \quad (1)$$

$$Precision = \frac{TP}{TP + FP} \quad (2)$$

$$Recall = \frac{TP}{TP + TN} \quad (3)$$

$$F1\text{-}Score = 2 \times \frac{Precision \times Recall}{Precision + Recall} \quad (4)$$

### C. Experiments

We started by following the data preprocessing step and followed the processes highlighted in section III.B. Following this, the hybrid feature selection algorithm was coded. The 81 features of the CICIDS 2017 dataset were passed into this hybrid selection algorithm model, which resulted in the selection of 32 key features. These 32 features were passed into 5 machine learning models. The results of these models were used to compare the efficiency of the combination of the hybrid feature selection algorithm and each of the five machine learning models to other baseline models. Before training the model, the dataset was split into 70% training set and 30% test set.

#### C.A.1. CIC IoT 2023 [19] dataset

The 18 features selected from this dataset using the proposed feature selection algorithm were used to train the Random Forest [13], Decision Trees [16], and XGBoost [40] model. The models were trained with default configurations. For each of the model, the accuracy and recall were measured, which were used to find out the best model out of the 3 models. Each of the models were trained on 583,250 benign traffic flows, and 24,201,239 DdoS flows as shown in Figure 2.

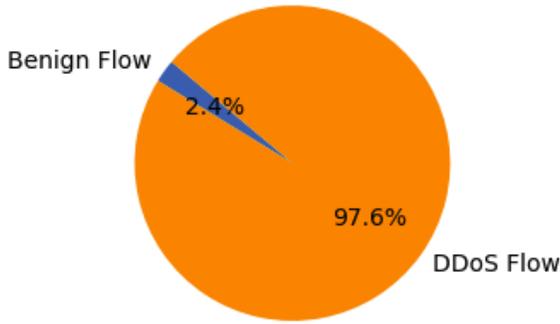

Fig. 2. CIC IoT 2023 Pie chart

*C.A.2. CIC IDS 2017[18]*

The 32 features selected from this dataset using the proposed feature selection algorithm were used to train the Random Forest [13], Decision Trees [16], Linear Support Vector Machines [40], XGBoost [41], and K-Nearest Neighbors [42] models. The models were trained with default configurations. For each of the models, the accuracy, recall, precision, and F1-Score was measured to compare the performance of these models against each other. Along with this, the time taken to train the model also formed a part of this experiment. Along with that, cross-validation [28] was applied to each model with 5 folds. Each of the models were trained on a dataset of 626,696 benign packets, and 128,025 DdoS packets. This is shown in Figure 3.

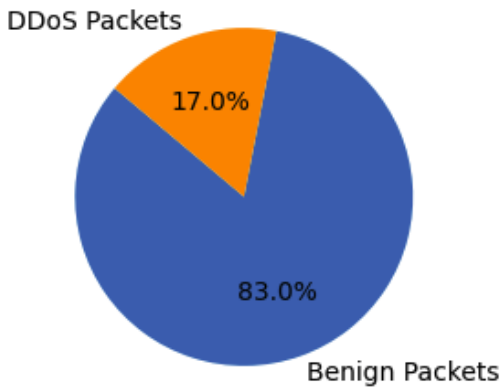

Fig. 3. CIC IDS 2017 [18] Pie chart

V. RESULTS AND COMPARISON

We found the evaluation metrics on the combination of the proposed hybrid feature selection algorithm with each of the following models on both the datasets:
1) Random Forest [13]
2) Decision Trees [16]
3) Linear Support Vector Machines [40]
4) XGBoost [41]
5) K-Nearest Neighbors [42]

*A. CIC IDS 2017 [18]*

*V.A.1 Random Forest [13] Model*

The accuracy, weighted average precision, weighted average F1-score, and the weighted average recall of the model turned out to be 99.989%. The amount of time it took to train this model was 52.94 seconds. This is shown in Table I. Cross-validation [28] was applied to the model with 5 folds. The results of cross-validation [28] on this model indicates that the model performed close to 99.99% in each of the 5 folds. This is shown in Table II.

*V.A.2. Decision Trees [16] Model*

The accuracy, weighted average precision, weighted average F1-score, and the weighted average recall of the model turned out to be 99.986%. The total time taken to train this model was 5.13 seconds. This is shown in Table I. Cross-validation [28] was applied to the model with 5 folds. The results of cross-validation [28] on this model indicates that the model performed close to 99.99% in each of the 5 folds. This is shown in Table II.

*V.A.3. Linear Support Vector Machines [40] Model*

The accuracy, weighted average precision, weighted average F1-score, and the weighted average recall of the model turned out to be 88.516%, 92.962%, 89.523%, and 88.516% respectively. The total time taken to train the Linear SVM [18] model was 139.37 seconds. This is shown in Table I. Cross-validation [28] was applied to the model with 5 folds. The results of cross-validation [28] on this model indicates that the model performance ranged from 90% to 97% in each of the 5 folds. This is shown in Table II.

*V.A.4. XGBoost [41] Model*

The accuracy, weighted average precision, weighted average F1-score, and the weighted average recall of the model turned out to be 99.993%. The amount of time taken to train this model was 1.95 seconds. This is shown in Table I.This is shown in Table I.Cross-validation [28] was applied to the model with 5 folds. The results of cross-validation [28] on this model indicates that the model performed close to 99.99% in each of the 5 folds. This is shown in Table II.

*V.A.5. K-Nearest Neighbors [42] Model*

The accuracy, weighted average precision, weighted average F1-score, and the weighted average recall of the model turned out to be 99.867%. The amount of time taken to train the model was 78.82 seconds. This is shown in Table I. Cross-validation [28] was applied to the model with 5 folds. The results of cross-validation [28] indicates that the model performance ranged from 99.8% to 99.9% in each of the 5 folds. This is shown in Table II.

### V.A.6 Discussion of Results

Overall, it was found that the XGBoost [41] model trained on the 32 features selected by the proposed hybrid feature selection algorithm got the highest accuracy, precision, recall, and F1-score of 99.993 %. In terms of time taken, the XGBoost [41] model took the least amount of time compared to the other models.

The model proposed by Bouke et al. [39] got an accuracy, precision, recall and Fscore of 98%. The model proposed by Chanu et al. [43] got an accuracy and precision of 98.8% and 99% respectively. The model proposed by Kshirsagar & Kumar [44] got an accuracy of 99.6238 %, a F1-Score of 99.9887 % with a model training time of 11.08 seconds. The maximum accuracy of the model proposed by Aysa et al. [14] was 99.7 %. My proposed model outperformed all of these models by attaning an accuracy, precision, recall, and F1-Score of 99.993%.

TABLE I: RESULTS FOR CIC-IDS 2017 DATASET

| Metrics | Random Forest | Decision Trees | **XGBoost** | Linear SVM | KNN |
|---|---|---|---|---|---|
| Accuracy | 99.989 % | 99.986 % | **99.993 %** | 88.516 % | 99.867 % |
| Precision | 99.989 % | 99.986 % | **99.993 %** | 92.962 % | 99.867 % |
| Recall | 99.989 % | 99.986 % | **99.993 %** | 88.516 % | 99.867 % |
| F1-Score | 99.989 % | 99.986 % | **99.993 %** | 83.530 % | 99.867 % |
| Training Time. | 52.94 seconds | 5.13 seconds | **1.95 seconds** | 139.37 seconds | 78.82 seconds |

TABLE III: RESULTS OF CROSS-VALIDATION ON ALL THE MODELS

| Model | Cross Validation Fold 1 (%) | Cross Validation Fold 2 (%) | Cross Validation Fold 3 (%) | Cross Validation Fold 4 (%) | Cross Validation Fold 5 (%) |
|---|---|---|---|---|---|
| Random Forest | 99.988 | 99.990 | 99.994 | 99.989 | 99.989 |
| Decision Trees | 99.988 | 99.988 | 99.987 | 99.993 | 99.990 |
| Linear SVM | 96.791 | 96.903 | 99.634 | 90.770 | 91.757 |
| **XGBoost** | **99.994** | **99.997** | **99.994** | **99.996** | **99.994** |
| KNN | 99.846 | 99.862 | 99.843 | 99.874 | 99.855 |

## B. CIC IoT 2023[19]

### V.B.1 Random Forest [13] Model

The accuracy for this model turned out to be 97.631 %. The precision turned out to be 95.45 %. The recall turned out to be 95.12 %. The F1-score turned out to be 95.28 %. The total amount of time taken to train this model was 3018.23 seconds. The mean squared error turned out to be 0.033. This is shown in Table III.

### V.B.2 XGBoost [41] Model

The accuracy for this model turned out to be 97.642 %. The total amount of time taken to train this model was 34.196 seconds. The precision turned out to be 95.33 %. The recall turned out to be 97.64 %. The F1-score turned out to be 96.47 %. The mean squared error turned out to be 0.023. This is shown in Table III.

### V.B.3 Decision Trees [17] Model

The accuracy for this model turned out to be 96.528 %. The total amount of time taken to train this model was 1185.16 seconds. The precision turned out to be 95.39 %. The recall turned out to be 96.52 %. The F1-score turned out to be 95.95 %. The mean squared error turned out to be 0.034. This is shown in Table III.

### V.B.4 Discussion

Overall, it was found that the XGBoost [41] model trained on the 18 features selected by the proposed hybrid feature selection algorithm got the best metrics compared to the other models trained. In terms of time taken, the XGBoost [41] model took the least amount of time compared to the other models. In terms of the mean squared error, XGBoost [41] model has the least mean squared error.

The best model proposed in [19] for a 2 class classification got an accuracy of 99.68%, recall of 96.51 %, precision of 96.53 %, and F1-score of 96.52%. Compared to this model, our proposed model has a better recall of 97.64 % and the percentages of accuracy, precision, and F1-score are extremely close to each other indicating that the values are comparable.

TABLE III: RESULTS OF THE MODELS APPLIED ON CIC IOT 2023 DATASET

| Metrics | Random Forest | Decision Trees | **XGBoost** |
|---|---|---|---|
| Accuracy | 97.631% | 96.528% | **97.642%** |

| Precision | 95.45% | 95.39% | **95.33%** |
|---|---|---|---|
| Recall | 95.12% | 96.52% | **97.64%** |
| F1-Score | 95.28% | 95.95% | **96.47%** |
| Training Time | 3018.23 seconds | 1185.16 seconds | **34.196 seconds** |
| Mean Squared Error | 0.033 | 0.034 | **0.023** |

*C. Explainability and Interpretability*

To increase the explainability and interpretability of the two proposed model, one for each dataset, we have done the analysis below. This analysis is critical to understand the reason for the predictions of the proposed models.

For the proposed model of the CIC IDS 2017 [18] dataset, it was found that certain features contributed more than the others to predict the results of the XGBoost [41] Model. These features are depicted in Figure 6. The reason that the model predicted the results with a 99.993 % accuracy is because some features like Init_Win_bytes_forward, Destination Port, Fwd Packets/s , Flow Duration, Init_Win_bytes_backward, Flow Packets/s, Fwd IAT Total, Fwd IAT Max, Bwd Packets, Fwd IAT Mean, Flow IAT Std, Bwd IAT Mac, Total length of Fwd Packets, and Flow IAT Max were some of the features out of the 32 selected features in the CIC IDS 2017 [18] dataset that provided the maximum contribution (in the order listed) towards predicting the results in the CIC IDS 2017 [18] dataset.

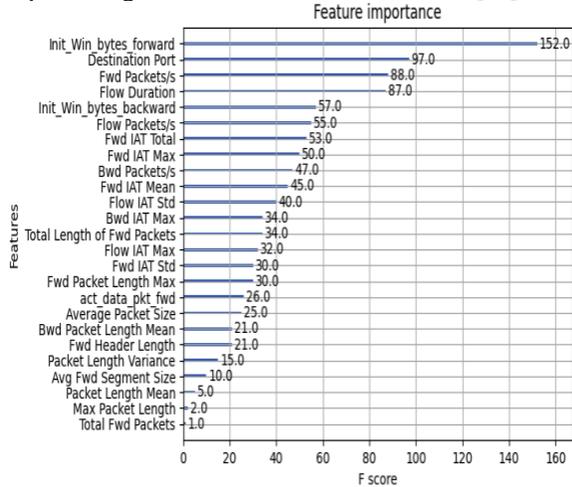

Fig. 3. Feature Importance CIC IDS 2017.

For the proposed model of the CIC IoT 2023 [19] dataset, it was found that certain features contributed more than the others to predict the results of the XGBoost [41] Model. These features are depicted in Figure 7. The reason that the model predicted the results with a 97.642 % accuracy is because some features like Variance, Protocol Tybe, ack_count, Number, syn_flag_number, HTTP, UDP, rst_flag_number, DHCP, ARP, SSH, ICMP, cwr_flag_number contributed more than the other selected features to predict the results of the model.

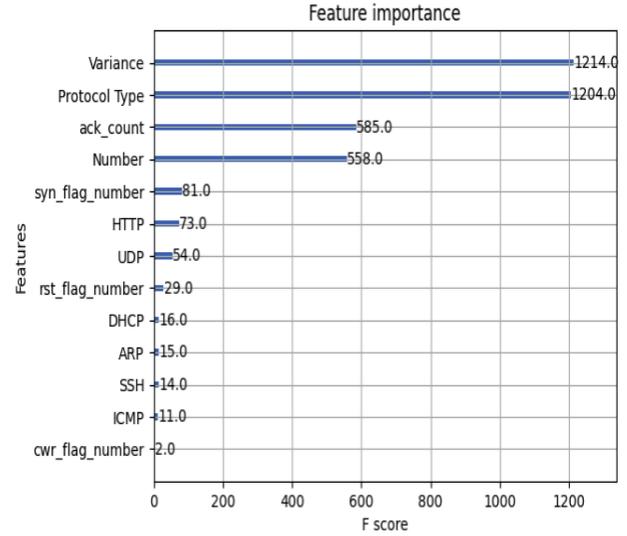

Fig. 4. Feature Importance CIC IoT 2023.

## VI. CONCLUSION AND FUTURE WORKS

In this paper, we have proposed a hybrid feature selection algorithm that selected 18 features in CIC IoT 2023 [19] dataset and 32 features in CIC IDS 2017 [18] dataset. The XGBoost [41] model trained with these features produced a superior accuracy of 99.993% on the CIC IDS 2017 [18] dataset and a superior recall of 97.64% on the CIC IoT 2023 [19] dataset. We also explained the predictions of both the models by explaining which features contributed the most towards the detection of DDoS attacks. In the future, we would like to explore the process of using more recent Explainable AI techniques to explain the findings of these models. Moreover, we would also like to explore the possibility of using deep learning [9] models combined with XGBoost [41] models to detect IoT DDoS attacks.

## VII. ACKNOWLEDGEMENTS

This work has been completed under the supervision of Dr. Rongxing Lu and supported as a research based course by the University of New Brunswick and the Canadian Institute of Cybersecurity.